\documentclass[nofootinbib,a4paper,aps,pra,showpacs,preprintnumbers,twocolumn]{revtex4-2}
\usepackage{cmap}
\usepackage{graphics,graphicx,epsfig}
\usepackage{epstopdf}
\usepackage[centertags]{amsmath}
\usepackage{amsfonts}
\usepackage{amssymb}
\usepackage[cp1251]{inputenc}
\usepackage[english]{babel}
\inputencoding{cp1251}
\usepackage{graphicx}
\usepackage{amsthm,color}
\usepackage{euscript}
\DeclareGraphicsExtensions{.pdf,.png,.jpg,.eps}
\begin{document}
\def\BY{\begin{eqnarray}}
\def\EY{\end{eqnarray}}
\def\L{\label}
\def\nn{\nonumber}
\def\ds{\displaystyle}
\def\o{\overline}
\def\({\left (}
\def\){\right )}
\def\[{\left [}
\def\]{\right]}
\def\<{\langle}
\def\>{\rangle}
\def\h{\hat}
\def\td{\tilde}
\def\r{\vec{r}}
\def\ro{\vec{\rho}}
\def\h{\hat}
\def\v{\vec}
\title{High-fidelity quantum gates for OAM qudits on quantum memory}
\author{Vashukevich E.A., Golubeva T.Yu., Golubev Yu.M.}

\affiliation{Saint Petersburg State University, Universitetskaya emb. 7/9, St. Petersburg, 199034 Russia}
\begin{abstract}
The application of high-dimensional quantum systems (qudits) in quantum computing and communications seems to be a promising avenue due to the possibility of increasing the amount of information encoded in one physical carrier. In this work, we propose a method for implementing single-qudit gates for qudits based on light modes with orbital angular momentum. Method for logical qudits encoding, which ensures the quasi-cyclicity of operations, is introduced. Based on the protocol for converting the orbital angular momentum of light in the Raman quantum memory scheme [Vashukevich E.A. et. al. PRA, 101, 033830 (2020)], we show that the considered gates provide an extremely high level of fidelity of single-qudit transformations. We also compare quantum gates' properties for systems of different dimensions and find the optimal conditions for carrying out transformations in the protocol under consideration.
\end{abstract}
\maketitle

\section{Introduction}

Several reasons explain the currently tremendous interest in high-dimensional quantum systems (qudits). Researchers are attracted by the possibility of increasing the information capacity of the channel -- the amount of information that can be encoded in one physical carrier, which turns out to be very useful in the problems of quantum communication \cite{Erhard2018a}. Significant advantages of qudits over qubits in quantum cryptography protocols have been demonstrated, where the security of the protocol appears to be the higher, the larger the dimension of the system \cite{sheridan2010security}. Many variants of various physical systems that could encode qudits have been proposed, for example, time-bin \cite{bechmann2000quantum}, the orbital angular momentum of light (OAM) \cite{Bent2015}, polarization multi-photon states \cite{bogdanov2004qutrit} and optical frequency comb \cite{Lukens17}.  Nevertheless, there are still blind spots in the problem of highly efficient manipulation of multidimensional quantum states.

The orbital angular momentum is an exciting resource for constructing a qudit since the OAM can take any integer values, which allows us to work in the Hilbert space of high dimension \cite{Allen92}. Laguerre-Gaussian (LG) modes with OAM show high stability and a relatively high decoherence time when propagating in a turbulent atmosphere \cite{LI2018}. Since LG modes are well localized in the spatial domain, several well-proven experimental techniques for generating \cite{slussarenko2011efficient, xiao2016generation}, separating and detecting such multimode radiation \cite{mirhosseini2013efficient, leach2002measuring, dai2015measuring} exist. Many methods of OAM manipulation based on phase holograms \cite{Heckenberg1992}, q-plates \cite{Karimi2009}, and a system of cylindrical lenses \cite{Beijersbergen1993} have also been proposed. However,  performing efficient mode conversions with different OAMs applying such optical elements requires a mode-specific change of the system parameters, which cannot satisfy quantum computing needs.
Although successful attempts to construct quantum logic gates over qudits based on OAM  have been made \cite{babazadeh2017}, the construction of universal gates that transform an arbitrary quantum state of light with OAM in a controlled manner with high fidelity and efficiency remains an open problem.

In the work \cite{vashukevich2020}, we demonstrated conversion of the OAM of light in the scheme of Raman quantum memory on cold atoms. We have shown that by varying the spatial profile of the driving field at the writing and read-out stages, the OAM of the quantum field can be changed by a certain amount in a wide range of values. Such a conversion scheme has relatively high efficiency (of the same order as the Raman memory protocols for light without OAM) and allows storing information simultaneously with the conversion.

In this work, we construct quantum single-qudit gates $ \h X_d^m $ based on the OAM transformation on a quantum memory cell for different dimensions of the qudits encoded in the OAM values. We show that with a specific encoding of the logical values of a qudit, it is possible to perform a cyclically closed operation $ \h X_d^m $. The proposed method is remarkable since it shows an extremely high fidelity, provides relatively high probabilities of success. In the last section, we compare systems of different dimensions in terms of both the properties of the gate $ \h X_d^m $ and the information capacity of the channel in an attempt to answer the question: "systems of what dimension are optimal for quantum computing?" in the presented protocol.

\section{Single-qudit multidimensional quantum gates}
Let us begin with recalling the general rules for constructing logical operations on high-dimensional systems. Following \cite{Lawrence2004} we introduce $ d $-dimensional logical operations on qudits $ \h X_d $ and $ \h Z_d $ using projection operators:
\BY
&&\h X_d=\sum\limits_{l=1}^{d}|l\oplus_d1\>\<l|,\;\;l\oplus_d1=l+1\;\;\hbox{mod}(d)\L{XD},\\
&&\h Z_d= \sum\limits_{l=1}^{d}|l\>\omega^l\<l|, \;\;\omega=\exp{2\pi i/d},\L{ZD}
\EY 
that is, the $ \h X_d $ gate performs modulo $ d $ addition of the value of the qudit with one, and the $ \h Z_d $ gate adds a relative phase to the members of the superposition. For $ d = 2 $, the matrices of such operators coincide with the Pauli matrices  $\h\sigma_x=\(\begin{smallmatrix}
		0&1\\1&0
	\end{smallmatrix}\),\h \sigma_z=\(\begin{smallmatrix}
		1&0\\0&-1
	\end{smallmatrix}\)$, and their action on an arbitrary qudit state $ |\psi\> $ can be described by the operation "NOT" and the operation of adding the phase $\exp{i \pi} =-1 $ to the second term of the superposition, respectively.
It is important to note that integer powers of mentioned above $ X_d, Z_d $ operators  should also be taken into account when considering $ d $ -dimensional operations on qudits:
\BY
&&\h X^m_d=\sum\limits_{l=1}^{d}|l\oplus_dm\>\<l|,\L{XDM}\\
&&\h Z^m_d=\sum\limits_{l=1}^{d}|l\>\omega^{ml}\<l|.\L{ZDM}
\EY 
The power $ m $ can range from $ 1 $ to $ d-1 $, since $ \h X_d^d =\h Z_d^d =I$. It is not difficult to write down the commutation relations for this operators, defining the Lie algebra $ su(d) $:
\BY
&&\[\h X_d^k,\h X_d^m\]=\[\h Z_d^k,\h Z_d^m\]=0,\L{COMM1}\\
&&\[\h X_d^k,\h Z_d^m\]=\sum\limits_{j=1}^d\(\omega^{mj}-\omega^{m(j\oplus_dk)}\)|j\oplus_dk\>\<j|.\L{COMM2}
\EY

Commutation relations for the Hermitian conjugate operators $ (\h X_d^k)^\dag, (\h Z_d^m)^\dag $ can be obtained from (\ref{COMM1}), (\ref{COMM2}) and following expressions:
\BY
&&(\h X_d^k)^\dag=\sum\limits_{j=1}^d|j\>\<j\oplus_dk|=\sum\limits_{j=1}^d|j \ominus_dk\>\<j|=\h X_d^{-k},\L{COMM12}\\
&&(\h Z_d^m)^\dag=\sum\limits_{j=1}^{d}|j\>(\omega^{mj})^*\<j|=\sum\limits_{j=1}^{d}|j\>\omega^{-mj}\<j|=\h Z_d^{-m}.\;\;\L{COMM22}
\EY
It is necessary to clarify the reason for considering only the operators $\h X_d, \h Z_d $ and their integer powers. As shown in the work \cite{babazadeh2017}, any unitary transformation over a single qudit $ \h U $ can be represented as a decomposition:
\BY
&&\h U=\sum\limits_{k=0}^{d-1}\sum\limits_{j=0}^{d-1}g_{j,k}\h X^j_d\h Z^k_d.\L{Unit}
\EY
For the sake of completeness, here we briefly recall the reasoning from \cite{babazadeh2017}, so consider the so-called Heisenberg-Weil operators:
\BY
&&\mathcal{\h D}(j,k)=\exp{\{i\frac{\pi kj}{2}\}}\h Z_d^j\h X_d^k\L{H-W}.
\EY
From these operators, one can compose a complete orthonormal set of operators in the Hilbert-Schmidt space $ \mathcal{L} (\mathcal{H}^d) $, which is the linear shell of the Hilbert space of dimension $ d $:
\BY&&\mathcal{\h Q}_{j,k}=\frac{1+i}{2}\mathcal{\h D}(j,k)+ \frac{1-i}{2}\mathcal{\h D}^\dag(j,k).\L{BAS}\EY

Any Hermitian operator $ \h A\in\mathcal{L}(\mathcal{H}^d) $ can be represented as a decomposition in a complete orthonormal set:
\BY
&&\h A=\sum\limits_{k=0}^{d-1}\sum\limits_{j=0}^{d-1}C_{j,k}\mathcal{\h Q}_{j,k}.\L{DEC}
\EY
Here $C_{j, k}$ are real decomposition coefficients. Any unitary transformation $ \h U $ can be written as a matrix exponent of the Hermitian operator:
\BY
&&\h U=\exp{\{iA\}}=\sum\limits_{n=0}^{\infty}\frac{i^n}{n!}\h A^n.\L{Unit2}\EY
Using the commutation relations (\ref{COMM1}), (\ref{COMM2}) and expressions (\ref{H-W})-(\ref{Unit2}), one can always write the unitary transformation matrix $\h U $ in the form (\ref{Unit}).

\section{Orbital angular momentum conversion}

From the previous section, one can see that the necessary universal set of single-qudit transformations consists of the operators $ \h X_d, \h Z_d $ and their integer powers, while the number of operations in the universal set increases with the dimension of the space. If we consider qudits based on orbital angular momentum, it turns out that the gate $ \h Z_d $ and its powers can be easily obtained using the Dove prism \cite{zhang2016}, so we will focus here on describing the protocol for implementing the gate $ \h X_d^m $ based on the method developed by us for transforming the OAM of the quantum field on the quantum memory cell \cite{vashukevich2020}.

Let us briefly recall the essence of the transformation of modes with OAM on a Raman quantum memory cell containing cold three-level atoms with $\Lambda $-configuration of energy levels. We have shown that it is possible to select the physical conditions of interaction that are consistent with the experimental possibilities, such that the modes of the quantum field with different OAM interact with the atomic ensemble independently of each other. If the driving field at the writing stage is a LG mode with OAM $ m $, and at the read-out stage it is a plane wave, the following expression describes the transformation of the quantum field in a complete memory cycle:
\BY
\h a^{out}_{l-m}(\td{t})&=&\chi_{l,m}\int\limits_0^{\td{T}} d\tilde{t}^\prime \h a^{in}_l(\tilde{t}^\prime) K(\td{t},\td{t}^{\prime}) +\h F_l.\L{68}
\EY
Under the action of the driving field, the orbital angular momentum of the output radiation changes. Here $ \h a^{out}_ {l-m}, \h a^{in}_l $ are the annihilation operators in LG modes with a certain OAM ($ {l-m} $ and $ l $) at the output and at the input of the memory cell, respectively, $ \h F_l $ are the operators of noise, which should be inevitably added during lossy transformations, $ K (\td{t}, \td{t}^{\prime}) $ is the kernel of a complete memory cycle. We assume that the memory protocol for spatial modes is optimized by choosing the effective cell length and temporal field profiles (a detailed analysis of the kernel is presented in \cite{Golubeva2012}), and the integral transformation with the kernel $ K (\td{t}, \td{t} ^ {\prime}) $ can be replaced with its unit eigenvalue. We want to highlight the coefficients in front of the integral $ \chi_{l, m} $ -- the mode overlap integrals normalized to the cross-sectional area of the LG mode. The coefficients $ \chi_{l, m} $ are determine the efficiency of the entire conversion. That is, we need to ensure a good overlap of spatial modes. To do this, we shift the waist of the driving and quantum fields by a certain amount $z_{S} $, assuming other parameters (waist width $ w_0 $ and Rayleigh range $ z_R $) of the beams to be the same. (see Fig. \ref{ZEFF}).
\begin{figure}
	\includegraphics[scale=0.3]{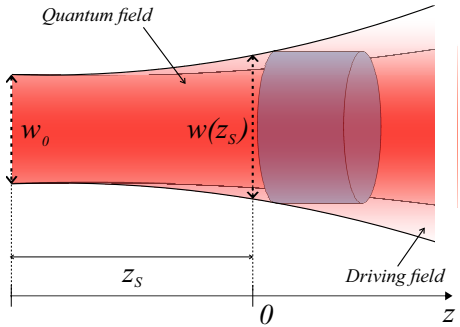}
	\caption{Schematic draw of the transformation geometry: the waist of the driving beam and the waist of the signal beam are shifted by $ z_S $, while the atomic ensemble is located in the waist of the signal field}\L{ZEFF}
\end{figure}
Such a parameter is introduced to follow the overlap of fields, and its variation allows us to control the efficiency of the transformation.

As one can see, in the process under consideration, the conversion of the orbital angular momentum of the quantum field occurs. Furthermore, if a field with a certain angular momentum $ l $ was written on the memory cell, then the OAM of the field at the output of the cell is equal to $ l-m $. Similarly, if the writing process is carried out by a plane wave, and the read-out is carried out by a field with OAM, then the orbital angular momentum of the field at the output will be equal to $ l + m $. The calculation shows that the conversion of any OAM up or down by 1 can be performed with high efficiency: about 0.9 for conversion with $ m=-1 $, and $ 0.6-0.8 $ for $m= + 1 $ \cite{vashukevich2020}. Since the kernel of a complete memory cycle $ K (\td{t}, \td{t}^{\prime}) $ does not depend on the index $ l $ of the signal field, we can transform the superposition of OAM states, where the OAM of all terms of the superposition increase or decrease by the same value $ m $.

\begin{figure*}[htb!]
	\center\includegraphics[scale=0.43]{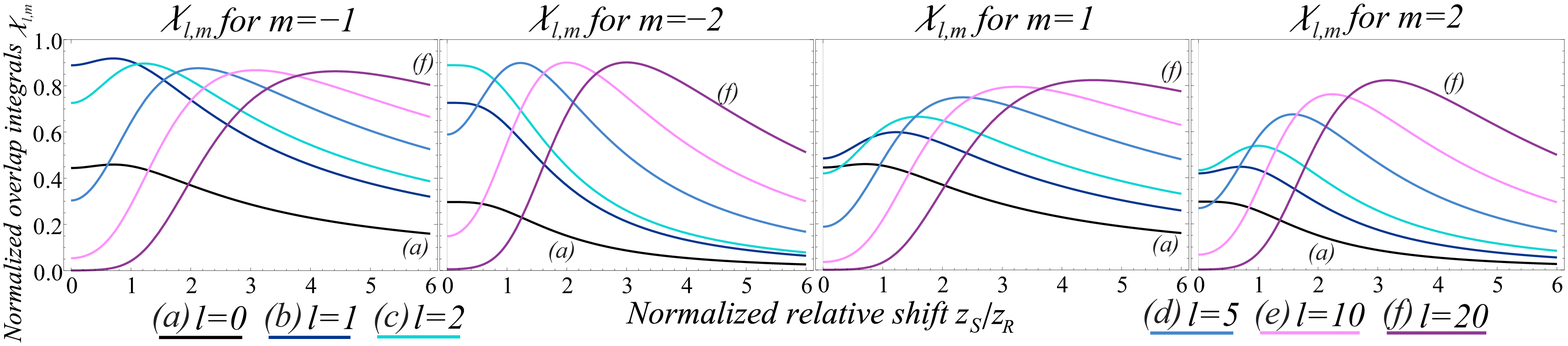}
	\caption{Coefficients $ \chi_{l, m} $ depending on the normalized relative shift of the beam waist $ z_S/z_R $ for different moments of the driving field $ m = -1, -2,1,2 $ (from left to right).}
	\L{FIGCHI2}
\end{figure*}

Fig. \ref{FIGCHI2} shows the dependence of the coefficients $ \chi_{l, m} $ for cases of writing by the driving field with OAM (from left to right) $ m = -1, -2,1,2 $ for different values of $ l $ depending on control parameter $ z_s / z_R $. Such a configuration of the fields makes it possible to change the OAM of the quantum field by the value $ m $ with the efficiency $ \chi_{l, m} $. We will denote by the lowercase letter $ \h x_d $ the lossy transformations, and by the capital letter $ \h X_d $ --the ideal one. Then the proposed configuration of the fields allows one to perform nonideal $ \h x^{-1}_d, \h x_d^{-2}, \h x_d^{}, \h x_d^{2} $, respectively. It can be noted that for the $ l = 0 $ the value of the coefficient $ \chi_{0, m} $ is rather small compared to the coefficients with nonzero $ l $. Since we are interested in performing transformations with high efficiency, we can conclude that using a state with $ l = 0 $ to build a state of a qudit is impractical. In addition, as it will be shown in the next section, transformations with $ m = 1,2 $ are noticeably loses to those with $ m = -1, -2 $ due to small values of the coefficients $\chi_{l, m} $ in the region of small values of $ l $.
\section{Orbital angular momentum qudits and quantum gates}
\subsection{Cyclicity of the gates and the concept of logical qudit}
As a starting point for describing the action of the OAM transformation on the states of the qudit, we consider an arbitrary multi-dimensional quantum state based on modes with OAM, which can be written in the following form:
\BY&&|\Psi\>= \sum\limits_{l=1}^{N}B_l|l\>.\L{ST}\EY
Here $ B_l $ are the decomposition coefficients, which obey the normalization condition $ \sum\limits_{l = 1}^{N}B^2_l = 1 $, $ l $ is the index associated with the orbital angular momentum. 

We consider a state with a certain OAM $ | l \> $ as a one-photon state generated as a result of the action of the creation operator $ \h a^\dag_l $ on the vacuum state:
\BY&&|l\>=\h a_l^\dag|0\>.\L{OAM}\EY
It should be noted here that we restrict the summation in the expression (\ref{ST}) within the range from $ 1 $ to $ N $ since consideration of the entire mode continuum with $ l \in (- \infty, \infty) $ is inconvenient for build a qudit for the reasons that we will discuss below.

Considering the $ N $ "physical" modes, let us single out several "logical" sets of dimensions $ d $ on this set:
\BY
&&l=\underbrace{1,2,...,d}_{\mbox{first logical set}},\underbrace{d+1,..2d}_{\mbox{second logical set}},...,\underbrace{...,N.}_{\mbox{N/d logical set}}
\EY
The state in which the index $ l $ varies within one logical set, we will assign to one qudit:
\BY
&&|\psi\>_n= \sum\limits_{l=(n-1)d+1}^{nd}C^n_l|l\>.\L{QU}
\EY
The index $ n $ numbers the "logical sets". The division into logical sets was made by us in order to cyclically close single-qudit operation $ \h X_d $:
\BY&& X_d|d\>_{physical}=|d+1\>_{physical}\equiv|1\>_{logical}\EY

If we consider as a qudit the whole state (\ref{ST}) (the entire continuum of modes with $ l \in (- \infty, \infty) $), then, obviously, no cyclically closed operation can be obtained. In our case, we can choose the "region" within the complete set of modes with OAM arbitrarily, depending on the experimental capabilities and computational needs.

Without loss of generality, let us suppose that $ N/d $ is an integer, then the state (\ref{ST}) can be rewritten as
\BY&&|\Psi\>= \sum\limits_{n=1}^{N/d}|\psi\>_n.\EY
We will further assume that different logical sets are indistinguishable. This assumption means that the state $ |l \> $ with OAM equal, for example, $ 1 $, $ d + 1 $ and $ 2d + 1 $ encode a logical unit. The states $ 2 $, $ d + 2 $ and $ 2d + 2 $ encode logical "2", and $ d $, $ 2d $ and $ 3d $ encode logical "$d $". Further we will consider only the state $ | \Psi \>_n $ as the state of the qudit and will omit the index $ n $  where possible.

Since proposed protocol for converting modes with OAM on a quantum memory cell \cite{vashukevich2020} use the notation of the creation and annihilation operators in modes with a specific OAM it seems reasonable to use the operator representation. The initial qudit state (\ref{QU}) can be generated from a vacuum as follows:
\BY
&&|\psi\>=\sum\limits_{l=1}^d C_l\h a^\dag_l|0\>=\h A^\dag|0\>.
\EY
Let the initial state of the qudit to be pure and described by the density matrix
$\h\rho_{in}$:
\BY
&&\h\rho_{in}=|\psi\>\<\psi|=\h A^\dag|0\>\<0|\h A.\L{pure}
\EY
The state of the qudit after applying the ideal gate $X_d^m $ can be written as
\BY
&&\h\rho_{ideal}=\h X_d^m|\psi\>\<\psi|\h X_d^{-m}=\h X_d^m\h A^\dag|0\>\<0|\h A\h X_d^{-m}=\nn\\
&&=\sum\limits_{l=1}^d C_l\h a^\dag_{l\oplus_d m}|0\>\<0|\sum\limits_{l=1}^d C^*_l\h a_{l\oplus_d m}.
\EY
The ideal transformation $ \h X_d^m $ "shifts" the mode index and does not change the coefficient $C_l$ -- the contribution to the superposition of each state with a certain OAM. Such a transformation has two essential properties: it occurs in a deterministic way (if we are talking about states with a few photons like (\ref{ST})) and "coherently" affects the state of the qudit, consistently transforming each member of the superposition. At the same time, we should ensure a "uniform" change in all physical states encoding the same logical qudit.  We will subsequently evaluate the imperfect transformation, which we will denote by $ \h x_d^m $, according to these criteria.

Moving on to the consideration of the nonideal transformation $ \h x_d^m $, from the expression (\ref{68}) we can conclude that the gate $ \h x_d^m $ has a probabilistic nature, since the coefficients $ \chi_{l, m} $, which determine the efficiency of the transformation of the OAM $ l $ by the value $ m $ is always less than one (see Fig. 2). In this case, having at the input a pure state of qudit (\ref{pure}), at the output we will obtain a statistical mixture described by the density matrix $ \h \rho_{out} $:
\BY
&&\h \rho_{out}=p_1\h\rho_1 + (1-p_1)\h\rho_2,\\
&&\h \rho_1=\frac{\h x_d^m\h\rho_{in}\h x_d^{-m}}{\hbox{Tr}\[\h x_d^m\h\rho_{in}\h x_d^{-m}\]}.
\EY
Here we especially highlighted the first term corresponding to the realization of the case when the entire state is successfully transformed, $ p_1 $ is the probability of the transformation, $ \h \rho_2 $ is the density matrix corresponding to all other outcomes (when not all modes were transformed). Renormalization to $ \hbox{Tr}\[\h x_d^m \h \rho_{in} \h x_d^{- m} \] $ is necessary to preserve the unit trace of the density matrix.

We will evaluate the conversion quality by the fidelity $ F $, calculated as follows:
\BY
F=\(\hbox{Tr}\sqrt{\sqrt{\h\rho_{ideal}}\h\rho_{1}\sqrt{\h\rho_{ideal}}}\)^2.
\EY
It should be noted that the expression above contains only the density matrix $ \h \rho_1 $, which corresponds to the outcome when a complete transformation of all modes occurred, and the fidelity written in this way shows only the degree of coherence of our transformation. The probability $ p_1 $ of the outcome we are interested in should be calculated individually. Based on the physical meaning of the density matrix $ \h \rho_1 $, we can write:
\BY
p_1=\prod\limits_{l=1}^{d}\chi^2_{l,m}.
\EY
Although the probabilities do not exceed 0.6, as shown in the next section, this problem can be eliminated in the experiment using the methods of postselection \cite {hofmann2002quantum, fattal2004quantum}.

Summarizing all of the above, we can emphasize several key aspects of the problem under consideration. The application of high-dimensional systems increases the channel's information capacity and the cryptographic security of the channel. However, as the dimension grows, the number of gates, which is necessary to construct an arbitrary unitary transformation (see (\ref{Unit})), also increases, and the probability of successful operation of the gate is reduced. Next, we will evaluate the influence of various factors and compare the results with a well-studied case of qubits.
\subsection{Computation of qutrit gates' properties}
This section will illustrate calculations of fidelity and probability for the case of qutrit, which, moreover, can be easily generalized to systems of higher dimensions.
For space dimension $ d = 3 $, the set of operations necessary for constructing arbitrary unitary transformation includes gates $ \h X_3, \h X_3^2, \h Z_3, \h Z_3^2 $. As mentioned earlier, the gates $ \h Z_d^m $ can be implemented using the Dove prism.

Based on the definition of (\ref{XDM}), we can conclude that the transformation $ \h X_3^2 $ is equivalent to the transformation $ \h X_3^{- 1} $, and $ \h X_3 $ is equivalent to $ \h X_3^{-2 } $ in the way of acting on an arbitrary state of qutrit, so in the further analysis, we will compare pairs of transformations with positive and negative powers.

Let us recall that not only the high probability of the gate success or the absolute values of the coefficients $ \chi_{l, m} $, is crucial to building the most optimal gate, but also the general "coherence" of the transformation, which can be formulated as the nearness of the values of the conversion coefficients for different members of the superposition. For example, if we work with the qutrit, which is encoded by the physical values of the OAM $ l-1, l, l + 1 $, then we must require the fulfilment of the following approximate equality:
\BY
\chi_{l-1, m}\approx\chi_{l, m}\approx\chi_{l+1, m}. \L{CHI}
\EY
Hence, a natural question arises: in what physical values of $ l $ should the qutrit be encoded so that the transformations of $ \h x_3^{m}, m =\pm1, \pm2 $ occur in the most optimal way? The calculation shows that as $ l $ grows, the conversion coefficients of different members of the superposition become closer to each other. That is, the condition (\ref {CHI}) is fulfilled the better, the bigger is $ l $. This result seems logical since $ \chi_{l, m} $ is accounted with geometric overlapping of the transverse spatial profiles of the modes (see \cite{vashukevich2020} for more details). However, the generation of qutrit states with large values of $ l $ is problematic from the experimental point of view. Therefore, we are especially interested in cases when relatively high values of probability and fidelity are attainable in the region of small $ l $.
\begin{figure}[htb!]
	\center\includegraphics[scale=0.55]{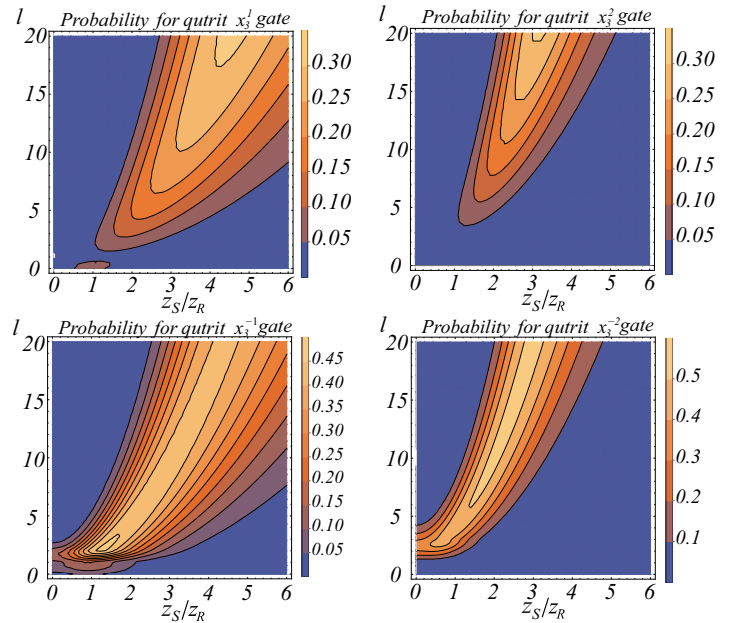}
	\caption{The probabilities of success of the qutrite gates $ x^{}_3, x^{2}_3 $ (top), $ x^ {-1}_3, x^{-2}_3 $ (bottom) depending on the values of $ l $ and of the parameter $ z_s / z_R $ that controls the geometry of the modes. As one can see, transformations with negative degrees have more attractive behaviour in the region of small OAM.}
	\L{Prob3}
\end{figure}
Fig. \ref{Prob3} shows the probabilities of success of gates $ x^{}_ 3, x^{2}_3 $ and $ x^{-1}_3, x^{-2}_3 $. Areas of high probabilities, as can be seen from the figure, show interesting behaviour. The regions expand with an increase of the control parameter $ z_s $ and an increase of the OAM $ l $. Moreover, with an increase in $ l $, they become more and more bent towards the vertical. This fact implies two remarkable properties of our transformation: firstly, for large $ l $ the probabilities will be high in a certain wide range of $ z_s $, and secondly, having an unknown state at the input of the system, we can fix some specific value of $ z_s $, which will provide an equally high probability of the gate success, practically (in the limit) independent of the OAM values of the input state when working in the region of large OAM values. However, since today the generation of light with a high OAM remains an experimentally laborious task, we will focus on the properties of transformations in the region of small $ l $. It can be noted that for $ x^{-1}_3, x^{-2}_3 $ the probabilities are higher than for the corresponding transformations with positive powers and, more importantly, high probabilities can be achieved in the region of small $ l $, that is a significant advantage over conversions $ x^{}_3, x^{2}_3 $.

One can also calculate fidelity values for these gates. The calculation shows that the proposed conversion method shows extremely high fidelity ($ F \geq 0.97 $) in a wide range of values of the control parameter $ z_s / z_R $ for various values of $ l $. Below we give specific values of the maximum fidelity (by varying the control parameter) for qutrit with values of $ l =  4,5,6$:
\BY
&F_{x_3}=1-8,9 \times10^{-6},
&F_{x^2_3}=1-26,2 \times10^{-6},\nn\\
&F_{x^{-1}_3}=1-5,5\times 10^{-6},
&F_{x^{-2}_3}=1-33,2\times 10^{-6}\nn.
\EY

Since the fidelity values turn out to be equally high for all transformations, we can conclude that it is most reasonable to perform transformations of the qutrit state with gates $ x^{- 1}_3, x^{- 2} _3 $.

Similar patterns can be seen for the transformation of quantum systems with a dimension greater than 3. Therefore, in the next section, we will compare the parameters of different gates for high-dimensional systems with the case of a qubit.
\subsection{Comparison of the gates for qudits with different dimension}
We have performed calculations of transformation properties for qudits of different dimensions (from $ d = 2 $ to $ 5 $) for the following states (the state numbers $ | l \> $ denote the physical value of the OAM projection):
\BY
&&|\psi\>^{d=2}= \frac{1}{\sqrt{2}}\(|4\>+|5\>\),\label{29}\\
&&|\psi\>^{d=3}= \frac{1}{\sqrt{3}}\(|4\>+|5\>+|6\>\),\\
&&|\psi\>^{d=4}= \frac{1}{\sqrt{4}}\(|4\>+|5\>+|6\>+|7\>\),\\
&&|\psi\>^{d=5}= \frac{1}{\sqrt{5}}\(|4\>+|5\>+|6\>+|7\>+|8\>\)\label{32}.
\EY
Such projections of the OAM $ l $ were chosen to encode the qudit state because for small values of $ l = 0, \pm1, \pm2 $, we will need to change the OAM of a state with $ l = 0 $  when performing transformations. These transformations occur with low efficiency due to weak overlap of the transverse profiles of the fundamental Gaussian mode with any LG modes with a nonzero OAM projection. The transition to the region of higher OAM values, as Fig. 3 shows, gives even higher values of probabilities, so the presented calculation can be considered as a lower bound.
\begin{table}[h]
	\caption{Comparison of gate $ x^{-1}_d $ parameters for qudits of different dimensions}
	\centering\begin{tabular}{|c|c|c|c|}
		\hline
		Qudit dim d & Probability P& Fidelity F & $\log_2d\times F\times P$ \\
		\hline
		$	2$ &$ 0,59 $&$ 1$ &$ 0,59 $\\
		\hline
		$	3 $&$ 0,44$ & $1-5,5\times 10^{-6}$ &$ 0,70 $\\
		\hline
		$4$ & $0,32$ &$ 1-37,4\times 10^{-6}$ &$0,65$ \\
		\hline
		$	5$ &$ 0,23$ & $1-80,6\times 10^{-6} $&$0,54$ \\
		\hline
	\end{tabular}
\end{table}
\begin{table}[h]
	\caption{Comparison of gate $x^{-2}_d$ parameters for qudits of different dimensions}
	\centering\begin{tabular}{|c|c|c|c|}
		\hline
		Qudit dim d & Probability P & Fidelity F & $\log_2d\times F\times P$ \\
		\hline
		$	2$ &$ 0,64 $&$ 1$ &$ 0,64 $\\
		\hline
		$	3 $&$ 0,49$ & $1-33,2\times 10^{-6}$ &$0,78 $\\
		\hline
		$4$ & $0,37$ &$ 1-218,0\times 10^{-6}$ &$0,74$ \\
		\hline
		$	5$ &$ 0,27$ & $1-457,9\times 10^{-6} $&$0,62$ \\
		\hline
	\end{tabular}
\end{table}

Table 1 shows a comparison of the properties of the $ x_d^{-1} $ transformation for systems of various dimensions described by states (\ref{29})-(\ref{32}). It can be noted that with an increase in the dimension (for $ d \geq3 $), both the probability of the gate success and the fidelity decrease, but the latter decreases very insignificantly. At the same time, in the last column of the table, the values of the product of fidelity, the probability, and the information capacity of the channel -- the number of bits of information encoded by one qudit of dimension $ d $, is shown. Thus, we wanted to emphasize that the loss in probabilities turns out to be not as significant as the gain in information capacity: indeed, it can be noted that for the case of qutrit and ququart, the value constructed in this way turns out to be larger than for other dimensions. Of particular interest are the results shown in Table 2, where the properties of the $ x_d ^ {- 2} $ gate are calculated. Its probability turns out to be greater than that of $x_d^{-1} $, which can be quite easily explained by purely geometric reasons of overlap of modes with close numbers.

Summarizing all of the above, we have shown that using a quantum memory scheme as a base for implementing single-qudit gates of various dimensions has significant advantages in extremely high fidelity and the ability to perform the storing of quantum state simultaneously with its transformation. The calculated probability of successful operation of the gates also remains high enough, and the scheme can be easily supplemented using postselection methods.
\section{Conclusion}

We have developed multidimensional quantum gates for qudits with OAM of different dimensions, based on the transformation of light modes with OAM in a quantum memory scheme. To ensure the cyclicity of the operations, we have identified several indistinguishable logical sets of dimension $ d $, encoding one qudit on the set of "physical" modes with OAM. Since the $\h Z_d$ transformation over light with OAM can be performed using Dove prisms, we focused on constructing the lossy gate $ \h x_d^m $ and comparing it with the ideal one $ \h X_d^m $. We carried out the comparison by calculating the probability of success and the fidelity of the transformation.

Calculations carried out for the particular case of the qudit dimension $ d = 3 $  showed that the considered gates reveal a relatively high probability of success (from 0.3 to 0.5). Moreover, for $ \h x^{-1}_3, \h x^{-2 }_3 $ high probability values are achieved in the region of small OAM $ l $, which seems to be a powerful argument in favor of working with transformations with negative powers of $ m $, since the experimental generation of light with high values of $ l $ has been currently facing with some difficulties. However, if efficient work with large $ l $ someday will be possible, then the proposed gates can provide equal probabilities when transforming an unknown state for a specific value of the control parameter $ z_s $ in a wide range of values of $ l $. This effect is achieved since the change in the regions of high probability with increasing $ z_s $ approaches the vertical line more and more. That is, the value of the coordinate $ z_s $, providing the highest probability, begins to depend weakly on the OAM values. A significant result is that all transformations provide an exceptionally high level of fidelity ($ F \geq97 $), which is also weakly dependent on the OAM.

We have compared the characteristics of the gates $ \h x^{- 1}_3, \h x^{-2}_3 $ for different dimensions of qudits with the trivial case of a qubit. The values of probability and fidelity were taken into account, along with the potential gain from information capacity increased with the dimension of a qudit. An estimation based on all these factors shows that working with qudits of dimensions $ d = 3 $ and $ d = 4 $, despite the lower probabilities, turns out to be preferable for performing quantum computations in the proposed protocol. A feature of the proposed transformations is that one can carry out computational procedures simultaneously with storing quantum-statistical properties of light in a memory cell.

This work was supported by the RFBR (grants 19-32-90059, 19-02-00204) and by the Foundation for the Advancement of Theoretical
Physics and Mathematics "BASIS" (grant 20-1-5-120-1).


\begin{thebibliography}{23}%
	\makeatletter
	\providecommand \@ifxundefined [1]{%
		\@ifx{#1\undefined}
	}%
	\providecommand \@ifnum [1]{%
		\ifnum #1\expandafter \@firstoftwo
		\else \expandafter \@secondoftwo
		\fi
	}%
	\providecommand \@ifx [1]{%
		\ifx #1\expandafter \@firstoftwo
		\else \expandafter \@secondoftwo
		\fi
	}%
	\providecommand \natexlab [1]{#1}%
	\providecommand \enquote  [1]{``#1''}%
	\providecommand \bibnamefont  [1]{#1}%
	\providecommand \bibfnamefont [1]{#1}%
	\providecommand \citenamefont [1]{#1}%
	\providecommand \href@noop [0]{\@secondoftwo}%
	\providecommand \href [0]{\begingroup \@sanitize@url \@href}%
	\providecommand \@href[1]{\@@startlink{#1}\@@href}%
	\providecommand \@@href[1]{\endgroup#1\@@endlink}%
	\providecommand \@sanitize@url [0]{\catcode `\\12\catcode `\$12\catcode
		`\&12\catcode `\#12\catcode `\^12\catcode `\_12\catcode `\%12\relax}%
	\providecommand \@@startlink[1]{}%
	\providecommand \@@endlink[0]{}%
	\providecommand \url  [0]{\begingroup\@sanitize@url \@url }%
	\providecommand \@url [1]{\endgroup\@href {#1}{\urlprefix }}%
	\providecommand \urlprefix  [0]{URL }%
	\providecommand \Eprint [0]{\href }%
	\providecommand \doibase [0]{https://doi.org/}%
	\providecommand \selectlanguage [0]{\@gobble}%
	\providecommand \bibinfo  [0]{\@secondoftwo}%
	\providecommand \bibfield  [0]{\@secondoftwo}%
	\providecommand \translation [1]{[#1]}%
	\providecommand \BibitemOpen [0]{}%
	\providecommand \bibitemStop [0]{}%
	\providecommand \bibitemNoStop [0]{.\EOS\space}%
	\providecommand \EOS [0]{\spacefactor3000\relax}%
	\providecommand \BibitemShut  [1]{\csname bibitem#1\endcsname}%
	\let\auto@bib@innerbib\@empty
	\bibitem [{\citenamefont {Erhard}\ \emph {et~al.}(2018)\citenamefont {Erhard},
		\citenamefont {Fickler}, \citenamefont {Krenn},\ and\ \citenamefont
		{Zeilinger}}]{Erhard2018a}%
	\BibitemOpen
	\bibfield  {author} {\bibinfo {author} {\bibfnamefont {M.}~\bibnamefont
			{Erhard}}, \bibinfo {author} {\bibfnamefont {R.}~\bibnamefont {Fickler}},
		\bibinfo {author} {\bibfnamefont {M.}~\bibnamefont {Krenn}},\ and\ \bibinfo
		{author} {\bibfnamefont {A.}~\bibnamefont {Zeilinger}},\ }\href
	{https://doi.org/10.1038/lsa.2017.146} {\bibfield  {journal} {\bibinfo
			{journal} {Light Sci. Appl.}\ }\textbf {\bibinfo {volume} {7}},\ \bibinfo
		{pages} {17111} (\bibinfo {year} {2018})},\ \Eprint
	{https://arxiv.org/abs/1708.06101} {1708.06101} \BibitemShut {NoStop}%
	\bibitem [{\citenamefont {Sheridan}\ and\ \citenamefont
		{Scarani}(2010)}]{sheridan2010security}%
	\BibitemOpen
	\bibfield  {author} {\bibinfo {author} {\bibfnamefont {L.}~\bibnamefont
			{Sheridan}}\ and\ \bibinfo {author} {\bibfnamefont {V.}~\bibnamefont
			{Scarani}},\ }\href@noop {} {\bibfield  {journal} {\bibinfo  {journal}
			{Physical Review A}\ }\textbf {\bibinfo {volume} {82}},\ \bibinfo {pages}
		{030301(R)} (\bibinfo {year} {2010})}\BibitemShut {NoStop}%
	\bibitem [{\citenamefont {Bechmann-Pasquinucci}\ and\ \citenamefont
		{Tittel}(2000)}]{bechmann2000quantum}%
	\BibitemOpen
	\bibfield  {author} {\bibinfo {author} {\bibfnamefont {H.}~\bibnamefont
			{Bechmann-Pasquinucci}}\ and\ \bibinfo {author} {\bibfnamefont
			{W.}~\bibnamefont {Tittel}},\ }\href@noop {} {\bibfield  {journal} {\bibinfo
			{journal} {Physical Review A}\ }\textbf {\bibinfo {volume} {61}},\ \bibinfo
		{pages} {062308} (\bibinfo {year} {2000})}\BibitemShut {NoStop}%
	\bibitem [{\citenamefont {Bent}\ \emph {et~al.}(2015)\citenamefont {Bent},
		\citenamefont {Qassim}, \citenamefont {Tahir}, \citenamefont {Sych},
		\citenamefont {Leuchs}, \citenamefont {S\'anchez-Soto}, \citenamefont
		{Karimi},\ and\ \citenamefont {Boyd}}]{Bent2015}%
	\BibitemOpen
	\bibfield  {author} {\bibinfo {author} {\bibfnamefont {N.}~\bibnamefont
			{Bent}}, \bibinfo {author} {\bibfnamefont {H.}~\bibnamefont {Qassim}},
		\bibinfo {author} {\bibfnamefont {A.~A.}\ \bibnamefont {Tahir}}, \bibinfo
		{author} {\bibfnamefont {D.}~\bibnamefont {Sych}}, \bibinfo {author}
		{\bibfnamefont {G.}~\bibnamefont {Leuchs}}, \bibinfo {author} {\bibfnamefont
			{L.~L.}\ \bibnamefont {S\'anchez-Soto}}, \bibinfo {author} {\bibfnamefont
			{E.}~\bibnamefont {Karimi}},\ and\ \bibinfo {author} {\bibfnamefont {R.~W.}\
			\bibnamefont {Boyd}},\ }\href {https://doi.org/10.1103/PhysRevX.5.041006}
	{\bibfield  {journal} {\bibinfo  {journal} {Phys. Rev. X}\ }\textbf {\bibinfo
			{volume} {5}},\ \bibinfo {pages} {041006} (\bibinfo {year}
		{2015})}\BibitemShut {NoStop}%
	\bibitem [{\citenamefont {Bogdanov}\ \emph {et~al.}(2004)\citenamefont
		{Bogdanov}, \citenamefont {Chekhova}, \citenamefont {Kulik}, \citenamefont
		{Maslennikov}, \citenamefont {Zhukov}, \citenamefont {Oh},\ and\
		\citenamefont {Tey}}]{bogdanov2004qutrit}%
	\BibitemOpen
	\bibfield  {author} {\bibinfo {author} {\bibfnamefont {Y.~I.}\ \bibnamefont
			{Bogdanov}}, \bibinfo {author} {\bibfnamefont {M.~V.}\ \bibnamefont
			{Chekhova}}, \bibinfo {author} {\bibfnamefont {S.~P.}\ \bibnamefont {Kulik}},
		\bibinfo {author} {\bibfnamefont {G.~A.}\ \bibnamefont {Maslennikov}},
		\bibinfo {author} {\bibfnamefont {A.~A.}\ \bibnamefont {Zhukov}}, \bibinfo
		{author} {\bibfnamefont {C.~H.}\ \bibnamefont {Oh}},\ and\ \bibinfo {author}
		{\bibfnamefont {M.~K.}\ \bibnamefont {Tey}},\ }\href@noop {} {\bibfield
		{journal} {\bibinfo  {journal} {Physical review letters}\ }\textbf {\bibinfo
			{volume} {93}},\ \bibinfo {pages} {230503} (\bibinfo {year}
		{2004})}\BibitemShut {NoStop}%
	\bibitem [{\citenamefont {Lukens}\ and\ \citenamefont
		{Lougovski}(2017)}]{Lukens17}%
	\BibitemOpen
	\bibfield  {author} {\bibinfo {author} {\bibfnamefont {J.~M.}\ \bibnamefont
			{Lukens}}\ and\ \bibinfo {author} {\bibfnamefont {P.}~\bibnamefont
			{Lougovski}},\ }\href {https://doi.org/10.1364/OPTICA.4.000008} {\bibfield
		{journal} {\bibinfo  {journal} {Optica}\ }\textbf {\bibinfo {volume} {4}},\
		\bibinfo {pages} {8} (\bibinfo {year} {2017})}\BibitemShut {NoStop}%
	\bibitem [{\citenamefont {Allen}\ \emph {et~al.}(1992)\citenamefont {Allen},
		\citenamefont {Beijersbergen}, \citenamefont {Spreeuw},\ and\ \citenamefont
		{Woerdman}}]{Allen92}%
	\BibitemOpen
	\bibfield  {author} {\bibinfo {author} {\bibfnamefont {L.}~\bibnamefont
			{Allen}}, \bibinfo {author} {\bibfnamefont {M.~W.}\ \bibnamefont
			{Beijersbergen}}, \bibinfo {author} {\bibfnamefont {R.~J.~C.}\ \bibnamefont
			{Spreeuw}},\ and\ \bibinfo {author} {\bibfnamefont {J.~P.}\ \bibnamefont
			{Woerdman}},\ }\href {https://doi.org/10.1103/PhysRevA.45.8185} {\bibfield
		{journal} {\bibinfo  {journal} {Phys. Rev. A}\ }\textbf {\bibinfo {volume}
			{45}},\ \bibinfo {pages} {8185} (\bibinfo {year} {1992})}\BibitemShut
	{NoStop}%
	\bibitem [{\citenamefont {Li}\ \emph {et~al.}(2018)\citenamefont {Li},
		\citenamefont {Chen}, \citenamefont {Gao}, \citenamefont {Willner},\ and\
		\citenamefont {Wang}}]{LI2018}%
	\BibitemOpen
	\bibfield  {author} {\bibinfo {author} {\bibfnamefont {S.}~\bibnamefont
			{Li}}, \bibinfo {author} {\bibfnamefont {S.}~\bibnamefont {Chen}}, \bibinfo
		{author} {\bibfnamefont {C.}~\bibnamefont {Gao}}, \bibinfo {author}
		{\bibfnamefont {A.~E.}\ \bibnamefont {Willner}},\ and\ \bibinfo {author}
		{\bibfnamefont {J.}~\bibnamefont {Wang}},\ }\href
	{https://doi.org/https://doi.org/10.1016/j.optcom.2017.09.034} {\bibfield
		{journal} {\bibinfo  {journal} {Optics Communications}\ }\textbf {\bibinfo
			{volume} {408}},\ \bibinfo {pages} {68} (\bibinfo {year} {2018})},\ \bibinfo
	{note} {optical Communications Exploiting the Space Domain}\BibitemShut
	{NoStop}%
	\bibitem [{\citenamefont {Slussarenko}\ \emph {et~al.}(2011)\citenamefont
		{Slussarenko}, \citenamefont {Karimi}, \citenamefont {Piccirillo},
		\citenamefont {Marrucci},\ and\ \citenamefont
		{Santamato}}]{slussarenko2011efficient}%
	\BibitemOpen
	\bibfield  {author} {\bibinfo {author} {\bibfnamefont {S.}~\bibnamefont
			{Slussarenko}}, \bibinfo {author} {\bibfnamefont {E.}~\bibnamefont {Karimi}},
		\bibinfo {author} {\bibfnamefont {B.}~\bibnamefont {Piccirillo}}, \bibinfo
		{author} {\bibfnamefont {L.}~\bibnamefont {Marrucci}},\ and\ \bibinfo
		{author} {\bibfnamefont {E.}~\bibnamefont {Santamato}},\ }\href@noop {}
	{\bibfield  {journal} {\bibinfo  {journal} {JOSA A}\ }\textbf {\bibinfo
			{volume} {28}},\ \bibinfo {pages} {61} (\bibinfo {year} {2011})}\BibitemShut
	{NoStop}%
	\bibitem [{\citenamefont {Xiao}\ \emph {et~al.}(2016)\citenamefont {Xiao},
		\citenamefont {Klitis}, \citenamefont {Li}, \citenamefont {Chen},
		\citenamefont {Cai}, \citenamefont {Sorel},\ and\ \citenamefont
		{Yu}}]{xiao2016generation}%
	\BibitemOpen
	\bibfield  {author} {\bibinfo {author} {\bibfnamefont {Q.}~\bibnamefont
			{Xiao}}, \bibinfo {author} {\bibfnamefont {C.}~\bibnamefont {Klitis}},
		\bibinfo {author} {\bibfnamefont {S.}~\bibnamefont {Li}}, \bibinfo {author}
		{\bibfnamefont {Y.}~\bibnamefont {Chen}}, \bibinfo {author} {\bibfnamefont
			{X.}~\bibnamefont {Cai}}, \bibinfo {author} {\bibfnamefont {M.}~\bibnamefont
			{Sorel}},\ and\ \bibinfo {author} {\bibfnamefont {S.}~\bibnamefont {Yu}},\
	}\href@noop {} {\bibfield  {journal} {\bibinfo  {journal} {Optics express}\
		}\textbf {\bibinfo {volume} {24}},\ \bibinfo {pages} {3168} (\bibinfo {year}
		{2016})}\BibitemShut {NoStop}%
	\bibitem [{\citenamefont {Mirhosseini}\ \emph {et~al.}(2013)\citenamefont
		{Mirhosseini}, \citenamefont {Malik}, \citenamefont {Shi},\ and\
		\citenamefont {Boyd}}]{mirhosseini2013efficient}%
	\BibitemOpen
	\bibfield  {author} {\bibinfo {author} {\bibfnamefont {M.}~\bibnamefont
			{Mirhosseini}}, \bibinfo {author} {\bibfnamefont {M.}~\bibnamefont {Malik}},
		\bibinfo {author} {\bibfnamefont {Z.}~\bibnamefont {Shi}},\ and\ \bibinfo
		{author} {\bibfnamefont {R.~W.}\ \bibnamefont {Boyd}},\ }\href@noop {}
	{\bibfield  {journal} {\bibinfo  {journal} {Nature communications}\ }\textbf
		{\bibinfo {volume} {4}},\ \bibinfo {pages} {1} (\bibinfo {year}
		{2013})}\BibitemShut {NoStop}%
	\bibitem [{\citenamefont {Leach}\ \emph {et~al.}(2002)\citenamefont {Leach},
		\citenamefont {Padgett}, \citenamefont {Barnett}, \citenamefont
		{Franke-Arnold},\ and\ \citenamefont {Courtial}}]{leach2002measuring}%
	\BibitemOpen
	\bibfield  {author} {\bibinfo {author} {\bibfnamefont {J.}~\bibnamefont
			{Leach}}, \bibinfo {author} {\bibfnamefont {M.~J.}\ \bibnamefont {Padgett}},
		\bibinfo {author} {\bibfnamefont {S.~M.}\ \bibnamefont {Barnett}}, \bibinfo
		{author} {\bibfnamefont {S.}~\bibnamefont {Franke-Arnold}},\ and\ \bibinfo
		{author} {\bibfnamefont {J.}~\bibnamefont {Courtial}},\ }\href@noop {}
	{\bibfield  {journal} {\bibinfo  {journal} {Physical review letters}\
		}\textbf {\bibinfo {volume} {88}},\ \bibinfo {pages} {257901} (\bibinfo
		{year} {2002})}\BibitemShut {NoStop}%
	\bibitem [{\citenamefont {Dai}\ \emph {et~al.}(2015)\citenamefont {Dai},
		\citenamefont {Gao}, \citenamefont {Zhong}, \citenamefont {Na},\ and\
		\citenamefont {Wang}}]{dai2015measuring}%
	\BibitemOpen
	\bibfield  {author} {\bibinfo {author} {\bibfnamefont {K.}~\bibnamefont
			{Dai}}, \bibinfo {author} {\bibfnamefont {C.}~\bibnamefont {Gao}}, \bibinfo
		{author} {\bibfnamefont {L.}~\bibnamefont {Zhong}}, \bibinfo {author}
		{\bibfnamefont {Q.}~\bibnamefont {Na}},\ and\ \bibinfo {author}
		{\bibfnamefont {Q.}~\bibnamefont {Wang}},\ }\href@noop {} {\bibfield
		{journal} {\bibinfo  {journal} {Optics letters}\ }\textbf {\bibinfo {volume}
			{40}},\ \bibinfo {pages} {562} (\bibinfo {year} {2015})}\BibitemShut
	{NoStop}%
	\bibitem [{\citenamefont {Heckenberg}\ \emph {et~al.}(1992)\citenamefont
		{Heckenberg}, \citenamefont {McDuff}, \citenamefont {Smith},\ and\
		\citenamefont {White}}]{Heckenberg1992}%
	\BibitemOpen
	\bibfield  {author} {\bibinfo {author} {\bibfnamefont {N.~R.}\ \bibnamefont
			{Heckenberg}}, \bibinfo {author} {\bibfnamefont {R.}~\bibnamefont {McDuff}},
		\bibinfo {author} {\bibfnamefont {C.~P.}\ \bibnamefont {Smith}},\ and\
		\bibinfo {author} {\bibfnamefont {A.~G.}\ \bibnamefont {White}},\ }\href
	{https://doi.org/10.1364/OL.17.000221} {\bibfield  {journal} {\bibinfo
			{journal} {Opt. Lett.}\ }\textbf {\bibinfo {volume} {17}},\ \bibinfo {pages}
		{221} (\bibinfo {year} {1992})}\BibitemShut {NoStop}%
	\bibitem [{\citenamefont {Karimi}\ \emph {et~al.}(2009)\citenamefont {Karimi},
		\citenamefont {Piccirillo}, \citenamefont {Nagali}, \citenamefont
		{Marrucci},\ and\ \citenamefont {Santamato}}]{Karimi2009}%
	\BibitemOpen
	\bibfield  {author} {\bibinfo {author} {\bibfnamefont {E.}~\bibnamefont
			{Karimi}}, \bibinfo {author} {\bibfnamefont {B.}~\bibnamefont {Piccirillo}},
		\bibinfo {author} {\bibfnamefont {E.}~\bibnamefont {Nagali}}, \bibinfo
		{author} {\bibfnamefont {L.}~\bibnamefont {Marrucci}},\ and\ \bibinfo
		{author} {\bibfnamefont {E.}~\bibnamefont {Santamato}},\ }\href
	{https://doi.org/10.1063/1.3154549} {\bibfield  {journal} {\bibinfo
			{journal} {Applied Physics Letters}\ }\textbf {\bibinfo {volume} {94}},\
		\bibinfo {pages} {231124} (\bibinfo {year} {2009})},\ \Eprint
	{https://arxiv.org/abs/https://doi.org/10.1063/1.3154549}
	{https://doi.org/10.1063/1.3154549} \BibitemShut {NoStop}%
	\bibitem [{\citenamefont {Beijersbergen}\ \emph {et~al.}(1993)\citenamefont
		{Beijersbergen}, \citenamefont {Allen}, \citenamefont {van~der Veen},\ and\
		\citenamefont {Woerdman}}]{Beijersbergen1993}%
	\BibitemOpen
	\bibfield  {author} {\bibinfo {author} {\bibfnamefont {M.}~\bibnamefont
			{Beijersbergen}}, \bibinfo {author} {\bibfnamefont {L.}~\bibnamefont
			{Allen}}, \bibinfo {author} {\bibfnamefont {H.}~\bibnamefont {van~der
				Veen}},\ and\ \bibinfo {author} {\bibfnamefont {J.}~\bibnamefont
			{Woerdman}},\ }\href {https://doi.org/10.1016/0030-4018(93)90535-D}
	{\bibfield  {journal} {\bibinfo  {journal} {Opt. Commun.}\ }\textbf {\bibinfo
			{volume} {96}},\ \bibinfo {pages} {123} (\bibinfo {year} {1993})}\BibitemShut
	{NoStop}%
	\bibitem [{\citenamefont {Babazadeh}\ \emph {et~al.}(2017)\citenamefont
		{Babazadeh}, \citenamefont {Erhard}, \citenamefont {Wang}, \citenamefont
		{Malik}, \citenamefont {Nouroozi}, \citenamefont {Krenn},\ and\ \citenamefont
		{Zeilinger}}]{babazadeh2017}%
	\BibitemOpen
	\bibfield  {author} {\bibinfo {author} {\bibfnamefont {A.}~\bibnamefont
			{Babazadeh}}, \bibinfo {author} {\bibfnamefont {M.}~\bibnamefont {Erhard}},
		\bibinfo {author} {\bibfnamefont {F.}~\bibnamefont {Wang}}, \bibinfo {author}
		{\bibfnamefont {M.}~\bibnamefont {Malik}}, \bibinfo {author} {\bibfnamefont
			{R.}~\bibnamefont {Nouroozi}}, \bibinfo {author} {\bibfnamefont
			{M.}~\bibnamefont {Krenn}},\ and\ \bibinfo {author} {\bibfnamefont
			{A.}~\bibnamefont {Zeilinger}},\ }\href
	{https://doi.org/10.1103/PhysRevLett.119.180510} {\bibfield  {journal}
		{\bibinfo  {journal} {Phys. Rev. Lett.}\ }\textbf {\bibinfo {volume} {119}},\
		\bibinfo {pages} {180510} (\bibinfo {year} {2017})}\BibitemShut {NoStop}%
	\bibitem [{\citenamefont {Vashukevich}\ \emph {et~al.}(2020)\citenamefont
		{Vashukevich}, \citenamefont {Golubeva},\ and\ \citenamefont
		{Golubev}}]{vashukevich2020}%
	\BibitemOpen
	\bibfield  {author} {\bibinfo {author} {\bibfnamefont {E.~A.}\ \bibnamefont
			{Vashukevich}}, \bibinfo {author} {\bibfnamefont {T.~Y.}\ \bibnamefont
			{Golubeva}},\ and\ \bibinfo {author} {\bibfnamefont {Y.~M.}\ \bibnamefont
			{Golubev}},\ }\href {https://doi.org/10.1103/PhysRevA.101.033830} {\bibfield
		{journal} {\bibinfo  {journal} {Phys. Rev. A}\ }\textbf {\bibinfo {volume}
			{101}},\ \bibinfo {pages} {033830} (\bibinfo {year} {2020})}\BibitemShut
	{NoStop}%
	\bibitem [{\citenamefont {Lawrence}(2004)}]{Lawrence2004}%
	\BibitemOpen
	\bibfield  {author} {\bibinfo {author} {\bibfnamefont {J.}~\bibnamefont
			{Lawrence}},\ }\href {https://doi.org/10.1103/PhysRevA.70.012302} {\bibfield
		{journal} {\bibinfo  {journal} {Phys. Rev. A}\ }\textbf {\bibinfo {volume}
			{70}},\ \bibinfo {pages} {012302} (\bibinfo {year} {2004})}\BibitemShut
	{NoStop}%
	\bibitem [{\citenamefont {Zhang}\ \emph {et~al.}(2016)\citenamefont {Zhang},
		\citenamefont {Roux}, \citenamefont {Konrad}, \citenamefont {Agnew},
		\citenamefont {Leach},\ and\ \citenamefont {Forbes}}]{zhang2016}%
	\BibitemOpen
	\bibfield  {author} {\bibinfo {author} {\bibfnamefont {Y.}~\bibnamefont
			{Zhang}}, \bibinfo {author} {\bibfnamefont {F.~S.}\ \bibnamefont {Roux}},
		\bibinfo {author} {\bibfnamefont {T.}~\bibnamefont {Konrad}}, \bibinfo
		{author} {\bibfnamefont {M.}~\bibnamefont {Agnew}}, \bibinfo {author}
		{\bibfnamefont {J.}~\bibnamefont {Leach}},\ and\ \bibinfo {author}
		{\bibfnamefont {A.}~\bibnamefont {Forbes}},\ }\href@noop {} {\bibfield
		{journal} {\bibinfo  {journal} {Science advances}\ }\textbf {\bibinfo
			{volume} {2}},\ \bibinfo {pages} {e1501165} (\bibinfo {year}
		{2016})}\BibitemShut {NoStop}%
	\bibitem [{\citenamefont {Golubeva}\ \emph {et~al.}(2012)\citenamefont
		{Golubeva}, \citenamefont {Golubev}, \citenamefont {Mishina}, \citenamefont
		{Bramati}, \citenamefont {Laurat},\ and\ \citenamefont
		{Giacobino}}]{Golubeva2012}%
	\BibitemOpen
	\bibfield  {author} {\bibinfo {author} {\bibfnamefont {T.~Y.}\ \bibnamefont
			{Golubeva}}, \bibinfo {author} {\bibfnamefont {Y.~M.}\ \bibnamefont
			{Golubev}}, \bibinfo {author} {\bibfnamefont {O.}~\bibnamefont {Mishina}},
		\bibinfo {author} {\bibfnamefont {A.}~\bibnamefont {Bramati}}, \bibinfo
		{author} {\bibfnamefont {J.}~\bibnamefont {Laurat}},\ and\ \bibinfo {author}
		{\bibfnamefont {E.}~\bibnamefont {Giacobino}},\ }\href
	{https://doi.org/10.1140/epjd/e2012-20723-3} {\bibfield  {journal} {\bibinfo
			{journal} {The European Physical Journal D}\ }\textbf {\bibinfo {volume}
			{66}},\ \bibinfo {pages} {275} (\bibinfo {year} {2012})}\BibitemShut
	{NoStop}%
	\bibitem [{\citenamefont {Hofmann}\ and\ \citenamefont
		{Takeuchi}(2002)}]{hofmann2002quantum}%
	\BibitemOpen
	\bibfield  {author} {\bibinfo {author} {\bibfnamefont {H.~F.}\ \bibnamefont
			{Hofmann}}\ and\ \bibinfo {author} {\bibfnamefont {S.}~\bibnamefont
			{Takeuchi}},\ }\href@noop {} {\bibfield  {journal} {\bibinfo  {journal}
			{Physical Review A}\ }\textbf {\bibinfo {volume} {66}},\ \bibinfo {pages}
		{024308} (\bibinfo {year} {2002})}\BibitemShut {NoStop}%
	\bibitem [{\citenamefont {Fattal}\ \emph {et~al.}(2004)\citenamefont {Fattal},
		\citenamefont {Diamanti}, \citenamefont {Inoue},\ and\ \citenamefont
		{Yamamoto}}]{fattal2004quantum}%
	\BibitemOpen
	\bibfield  {author} {\bibinfo {author} {\bibfnamefont {D.}~\bibnamefont
			{Fattal}}, \bibinfo {author} {\bibfnamefont {E.}~\bibnamefont {Diamanti}},
		\bibinfo {author} {\bibfnamefont {K.}~\bibnamefont {Inoue}},\ and\ \bibinfo
		{author} {\bibfnamefont {Y.}~\bibnamefont {Yamamoto}},\ }\href@noop {}
	{\bibfield  {journal} {\bibinfo  {journal} {Physical review letters}\
		}\textbf {\bibinfo {volume} {92}},\ \bibinfo {pages} {037904} (\bibinfo
		{year} {2004})}\BibitemShut {NoStop}%
\end{thebibliography}
\providecommand{\noopsort}[1]{}\providecommand{\singleletter}[1]{#1}%
%


%

\end{document}